\documentclass[preprint,nofootinbib]{revtex4-1}

\usepackage[utf8]{inputenc}
\usepackage{amsmath,amssymb,amsfonts,xcolor}

\newcommand{\ra}{\rightarrow}
\newcommand{\eps}{\epsilon}
\newcommand{\tE}{{E}}

\newcommand{\<}{\langle}
\renewcommand{\>}{\rangle}

\newtheorem{claim}{Claim}
\newtheorem{lemma}{Lemma}

\usepackage[colorlinks=true]{hyperref} 
\hypersetup{
    bookmarks=true,         
    unicode=false,          
    pdftoolbar=true,        
    pdfmenubar=true,        
    pdffitwindow=false,     
    pdfstartview={FitH},    
    pdfsubject={},   
    pdfcreator={},   
    pdfproducer={}, 
    pdfkeywords={} {} {}, 
    pdfnewwindow=true,      
    colorlinks=true,       
    linkcolor=magenta, 
    citecolor=blue,        
    filecolor=magenta,      
    urlcolor=blue           
}

\begin{document}

\title{Nernst and Ettingshausen effects in gapped quantum materials}
\author{Michael Levin}
\email{malevin@uchicago.edu}
\affiliation{University of Chicago}
\author{Anton Kapustin}
\email{kapustin@caltech.edu}
\affiliation{California Institute of Technology}
\author{Lev Spodyneiko}
\email{lionspo@caltech.edu}
\affiliation{California Institute of Technology}

\begin{abstract}
We investigate whether there could exist topological invariants of gapped 2D materials related to dissipationless thermoelectric transport at low temperatures. We give both macroscopic and microscopic arguments showing that thermoelectric transport coefficients vanish in the limit of zero temperature and thus topological invariants arise only from the electric Hall conductance and the thermal Hall conductance. Our arguments apply to systems with arbitrarily strong interactions. We also show that there is no analog of the Thouless pump for entropy. 
\end{abstract}

\maketitle

\section{Introduction}

One of the most striking features of topological phases of matter is their unusual transport properties. Famously, 2D materials with broken time-reversal symmetry and a bulk energy or mobility gap can exhibit electric Hall conductance and thermal Hall conductance which are not exponentially suppressed at low temperatures. This can be attributed to chiral gapless edge modes which carry both electric and energy  currents and are robust under arbitrary perturbations which do not close the gap. It is natural to ask whether other thermoelectric properties can exhibit similar anomalous behavior. In particular, can there be quantum Nernst and Ettingshausen effects, by analogy with the quantum  Hall and quantum thermal Hall effects? More generally, it is of interest to understand fundamental constraints on thermoelectric transport at low temperatures. 

In the linear regime, the Nernst and Ettingshausen effects in 2D materials are usually described in terms of two coefficients $\nu_{xy}$ and $\eta_{xy}$ which enter the phenomenological expressions for net electric and heat currents\footnote{If time-reversal symmetry is present, Onsager reciprocity implies  that $\eta_{xy}=T\nu_{yx}$, but in this paper we are interested in situations where time-reversal symmetry is  broken, either by an external magnetic field or spontaneously.}:
\begin{align}\label{elnet}
I^{el}_x & =-\sigma_{xy}(T) \Delta_y \mu -\nu_{xy}(T) \Delta_y T,\\ \label{heatnet}
I^{heat}_x & = -\eta_{xy}(T) \Delta_y \mu - \kappa_{xy}(T) \Delta_y T.
\end{align}
Here $\mu$ is the electrochemical potential (we are working in units where the electron charge $e$ is set to $1$), $T$ is the temperature, and we assumed that $\mu$ and $T$ depend only on the $y$-coordinate. 
The coefficients $\sigma_{xy}$ and $\kappa_{xy}$ are the Hall conductance and the thermal Hall conductance, respectively. The coefficients $\nu_{xy}$ and $\eta_{xy}$ are usually called transverse thermoelectric coefficients; we will call them the Nernst and Ettingshausen coefficients, respectively. 

The coefficients $\sigma_{xy}$ and $\nu_{xy}$ are dimensionless, while $\eta_{xy}$ and $\kappa_{xy}$ have the units of energy. Thus the quantum Nernst effect would mean that $\nu_{xy}(T)$ approaches a nonzero constant $\nu_{xy}(0)$ at $T=0$, while the quantum Ettingshausen effect would mean that $\eta_{xy}=b T+\ldots$, where $b$ is a nonzero constant and the dots denote terms which vanish faster that linearly (in a gapped material, presumably the dotted terms are exponentially suppressed). In a gapped material, there should be no appreciable bulk currents, and thus it should be possible to interpret the net currents (\ref{elnet}-\ref{heatnet}) in terms of edge currents. Indeed, it is easy to see that if $b=-\nu_{xy}(0)$, then one can attribute (\ref{elnet}-\ref{heatnet}) to chiral edge modes with equilibrium electric and energy currents
\begin{align}\label{eq:edge current 1}
I^{el}_{edge} &=-\sigma_{xy}(0) \mu-\nu_{xy}(0) T,\\\label{eq:edge current 2}
I^{E}_{edge} &=-\sigma_{xy}(0) \dfrac{\mu^2}2- \nu_{xy}(0) T\mu-\frac12 c T^2.
\end{align}
Here we also assumed that $\kappa_{xy}(T)=cT$ for some dimensionless constant $c$. F. Bloch's theorem \cite{Bohm,Watanabe} and its energy counterpart \cite{KapSpo} prohibit equilibrium currents in genuine 1D systems, but not for edges of 2D systems. 

For IQHE and FQHE systems, it is well-known that both $\sigma_{xy}(0)$ and $c$ are nonzero and are topological invariants. If $\nu_{xy}(0)$ were nonzero for some system, it would be a new topological invariant of gapped 2D materials. To see this, consider a strip of a 2D material infinitely extended in the $x$ direction but having a finite extent in the $y$ direction. Suppose  that the parameters of the Hamiltonian are slowly varying in the $y$ direction while maintaining a bulk gap. On the one hand, this system can be thought of as a 1D system, and therefore the net currents must vanish in equilibrium. Thus in equilibrium (that is, for constant $T$ and $\mu$) the edge currents must cancel between the two edges. On the other hand, the edge currents should be determined by the Hamiltonians near the respective edges. Thus $\sigma_{xy}(0),\nu_{xy}(0),$ and $c$ do not change under the variations of the Hamiltonian which do not close the bulk energy gap.

It is well-known that if the edges modes are non-interacting fermions, both $\nu_{xy}(T)$ and $\eta_{xy}(T)/T$ vanish at $T=0$.  This follows from two observations: (1) systems of free fermions have approximate particle-hole symmetry which becomes exact at $T=0$; (2) under the particle-hole symmetry, the temperature and the energy current are even, while the electrochemical  potential, the electric current, and the Nernst and Ettingshausen coefficients are odd. Thus a necessary condition for a nonzero $\nu_{xy}(0)$ are strong interactions.

In the case when the edge modes are described by a 1+1d Conformal Field Theory, it is well-known that non-zero values of $\sigma_{xy}(0)$ and $c$ are related to $U(1)$ and gravitational anomalies, respectively, which can be cancelled by Chern-Simons and gravitational Chern-Simons terms in the bulk. By analogy one might expect that a nonzero value of $\nu_{xy}(0)$ is related to topological terms in the bulk action involving both $U(1)$ and gravitational fields. Since such terms cannot be constructed, the conclusion seems to be that the quantum Nernst and Ettingshausen effects do not exist either. However, this argument relies on the assumption that every anomaly of the edge modes can be cancelled by a term in the bulk effective action. This is not obvious and in fact is not true in certain non-unitary conformal field theories (CFTs). Specifically, ghost CFTs have mixed $U(1)$-gravitational anomalies \cite{GreenSchwarzWitten}. In general, it is not understood which field theory anomalies can always be cancelled by bulk topological terms. 

In this paper we describe three arguments of varying generality showing that for gapped systems $\nu_{xy}(0)=\lim_{T\ra 0}\eta_{xy}(T)/T=0$ and thus the quantum Nernst and Ettingshausen effects do not exist. In Section II we set up the notation and recall the expressions for the entropy production rate and the entropy current. In Section III we deduce the vanishing of $\nu_{xy}(0)$ and $\lim_{T\ra 0}\eta_{xy}(T)/T$ from the Third Law of Thermodynamics. We also clarify the constraints imposed by the Third Law on the Seebeck and Peltier effects.  In Section IV we present another argument for the vanishing of $\nu_{xy}(0)$ and $\lim_{T\ra 0}\eta_{xy}(T)/T$, under the assumption that the edges are described by a {\it unitary} CFT.  
In Section V we present yet another argument for the vanishing of $\nu_{xy}(T)$ at low temperatures, based on a flux insertion thought experiment. This argument does not require a unitary CFT structure for edge excitations, but it does invoke some reasonable but hard-to-prove physical assumptions. 
As part of the latter argument, we also show that there is no analog of the Thouless pump for entropy. We discuss our results in Section VI.

The second-named author would like to thank Assa Auerbach for a discussion of St\v{r}eda formulas. The work of A. K. and L. S. was supported in part by the U.S.\ Department of Energy, Office of Science, Office of High Energy Physics, under Award Number DE-SC0011632. A.\ K.\ was also supported by the Simons Investigator Award.  M.\ L.\ was supported in part by the Simons Collaboration on Ultra-Quantum Matter, which is a grant from the Simons Foundation (651440).

\section{Generalities}

In this section we recall the definition of transport coefficients and some of their properties. The discussion applies both in 2D and 3D; in later sections we specialize to 2D materials.

In the hydrodynamic limit, one can expand the electric current density and the energy current density to first order in the electric field and the temperature gradient.\footnote{More precisely, these are ``transport'' currents. There are also ``magnetization'' currents which are present even in equilibrium.} For simplicity we will assume that the chemical potential is constant. For the electric current, this expansion has the form 
\begin{equation}
{\bf j}^N_k=\sigma_{km} {\bf \tE}_m -\nu_{km} \partial_m T,
\end{equation}
where ${\bf \tE}_k=-\partial_k\varphi-\partial_t {\bf A}_k$ and $\varphi$ is the electric potential. The conductivity tensor $\sigma_{km}$ and the thermoelectric tensor $\nu_{km}$ are functions of temperature only. Strictly speaking, this expansion applies in a non-equilibrium steady state (NESS). Thus we may assume $\nabla\times {\bf \tE}=\frac{\partial {\bf B}}{\partial t}=0$. For the energy current the expansion is 
\begin{equation}
 {\bf j}^E_k=\varphi {\bf j}^N_k+\eta_{km} {\bf \tE}_m -\kappa_{km} \partial_m T.
\end{equation}
Note that the first term on the r.h.s. is not invariant under a constant gauge transformation $\varphi(x)\mapsto\varphi(x)+c$. This is because the energy density operator also transforms under such gauge transformations, $\hat h({\bf r})\mapsto \hat h({\bf r})+c \hat\rho({\bf r})$, where $\hat\rho$ is the electric charge density. Since the current operators  are defined by the equations
\begin{equation}\label{def:currents}
i[\hat H,{\hat h}(x)]=-\nabla\cdot {\bf \hat j}^E(x),\quad i[\hat H,\hat\rho(x)]=-\nabla\cdot {\bf \hat j}^N(x),
\end{equation}
this requires the energy current operator to transform as well, ${\bf \hat j}^E\mapsto {\bf \hat j}^E+c {\bf \hat j}^N.$ The tensor $\nu_{km}$ describes the Seebeck and Nernst effects, while the tensor $\eta_{km}$ describes the Peltier and Ettingshausen effects.

Transport coefficients are constrained by Onsager's reciprocity relations \cite{LandauLifshits}:
\begin{equation}\label{onsager}
\sigma_{km}(T,{\bf B})=\sigma_{mk}(T,-{\bf B}),\quad \kappa_{km}(T,{\bf B})=\kappa_{mk}(T,-{\bf B}),\quad  \eta_{km}(T,{\bf B})=T\nu_{mk}(T,-{\bf B}).
\end{equation}
where we assumed for definiteness that time-reversal invariance is broken only by an external magnetic field ${\bf B}$. In general, Onsager reciprocity relates transport coefficients of a system and its time-reversal partner. Note that in a  time-reversal-invariant situation the tensors $\sigma$ and $\kappa$ are required to be symmetric, but $\nu$ and $\eta$ can have both a symmetric and an  anti-symmetric part. We will distinguish the symmetric and anti-symmetric components with superscripts $S$ and $A$. Thus $\sigma^S$ is the ordinary conductivity tensor, while $\sigma^A$ is the Hall conductivity tensor, etc.

One can compute the bulk entropy production rate per unit volume following \cite{LandauLifshits}. The rate of change of entropy density is
\begin{equation}
\frac{\partial s}{\partial t}=\frac{1}{T}\left(\frac{\partial h}{\partial t}-\varphi\frac{\partial \rho}{\partial t}\right)=\frac{1}{T}\left(- \nabla\cdot {\bf j}^E +\varphi \nabla\cdot {\bf j}^N\right).
\end{equation}
Computing the divergences of currents and taking into account $\nabla\times{\bf\tE}=0$ we get
\begin{equation}\label{entropyproduction}
\frac{\partial s}{\partial t}=\frac{1}{T}\sigma_{km} {\bf \tE}_k {\bf \tE}_m+\frac{1}{T^2}\kappa_{km} \partial_k T\partial_m T-\frac{1}{T^2}(T\nu_{km}+ \eta_{mk}) {\bf \tE}_k \partial_m T-\nabla\cdot {\bf j}^S_k,
\end{equation}
where the entropy current density is
\begin{equation}\label{entropycurrent}
{\bf j}^S_k=\frac{1}{T}\eta_{km} {\bf \tE}_m-\frac{1}{T}\kappa_{km} \partial_m T .
\end{equation}
Note that only the symmetric parts of the conductivity tensors $\sigma_{km}$ and $\kappa_{km}$ enter the expression for the entropy production rate. The anti-symmetric parts (the Hall conductance $\sigma^A$ and the thermal Hall conductance $\kappa^A$) drop out. The entropy current seems to depend on the whole tensor $\kappa$, but in fact one can replace $\kappa$ with $\kappa^S$, since this changes the current only by a divergence-free vector field. One can say that $\sigma^A$ and $\kappa^A$ describe non-dissipative phenomena. On the other hand, both symmetric and anti-symmetric parts of $\nu_{km}$ and $\eta_{km}$ contribute to dissipation. Note also that Onsager reciprocity relations ensure that the entropy production rate does not change if one switches the direction of the external magnetic field (or replaces the system with its time-reversal partner).

\section{Thermoelectric coefficients of gapped materials and the Third Law}
\label{sec:thirdlaw}

Consider a material with short-range interactions and either a bulk energy gap or a mobility gap.\footnote{The assumption about short-range interactions is made to exclude superconductors, where an energy gap arises only due to long-range Coulomb interactions.} If the temperature $T$ is well below the bulk gap, the entropy production rate per unit volume should be exponentially small. From (\ref{entropyproduction}) we see that both $\sigma^S(T)$ and $\kappa^S(T)$ are exponentially small, while $\eta_{mk}(T)\simeq -T\nu_{km}(T)$ up to exponentially small corrections.

Let us show that the Third Law of Thermodynamics implies $\lim_{T\ra 0}\eta^S(T)/T=\nu^S(0)=0$. One standard formulation of the Third Law (``the Nernst unattainability principle" \cite{NernstPrinciple}) states that it is impossible to lower the entropy of a body to its zero-temperature value within a finite time. The Nernst unattainability principle prohibits creating a {\it perpetuum mobile} of the third kind (a Carnot engine where one of the heat baths is at $T=0$).  Consider two heat baths at low $T$ connected by a  cylindrical ``bridge'' made of the material of interest. If one applies an electric field across the bridge, the entropy current density across the bridge is given by eq. (\ref{entropycurrent}). Since $\kappa^S(T)$ is exponentially small, the magnitude of the entropy current is determined by $\eta_{km}^S(T)\hat n_k\hat n_m/T $, where $\hat n_k$ is a unit vector in the direction of the cylinder axis. If $\lim_{T\ra 0}\eta^S(T)/T\neq 0$, a nonzero amount of entropy $\Delta S$ will be transported across the bridge per unit time even as $T$ approaches zero for some choice of how we orient the materials before wrapping it into the cylinder. If the excess entropy of the ``source'' heat bath over its zero-$T$ value is $S$, it would take a finite time $S/\Delta S$ to lower its entropy to its zero-$T$ value, in contradiction with the Nernst unattainability principle. Hence we must have $\lim_{T\ra 0}\eta^S(T)/T=0$. Then Onsager reciprocity (\ref{onsager}) implies $\nu^S(0)=0$. This argument is very robust and does not depend on the existence of the bulk  gap.  

To show that $\lim_{T\ra 0}\eta^A(T)/T=\nu^A(0)=0,$ we make use of another common formulation of the Third Law: ``the Nernst heat theorem". It states that the $T\ra 0$ limit of the entropy of a body is finite and independent of the parameters of the Hamiltonian. It does not follow from the Nernst unattainability principle without additional assumptions \cite{Landsberg} (for a recent discussion see \cite{Klimenko}). For example, the ideal Boltzmann gas violates the Nernst heat theorem because its entropy diverges as $T\ra 0$. Nevertheless, if one assumes that the $T\ra 0$ limit of the entropy is finite (``the Einstein principle") and that the specific heat is strictly positive, one can deduce the Nernst heat theorem from the Nernst unattainability principle \cite{Landsberg}.

Consider a piece of material shaped as a cylindrical shell with caps on both ends. The shell is in contact with a heat bath at temperature $T$. Suppose there is a current-carrying solenoid inside the cylinder so that the two caps are subject to a magnetic field $\pm B$. We also assume that the material is not completely homogeneous and interpolates between a trivial insulator at one end and a gapped  material of interest at the other end. At strictly zero $T$ this might lead to a divergent correlation length at the interface between the two materials. However, at $T>0$ one expects such an interpolation to be possible while maintaining a finite correlation length throughout. For example, if the interface is described by a 1+1d CFT, the correlation length is of order $1/T$. This ensures that a hydrodynamic description is possible. 

The main ingredient in the argument is the thermodynamic formula for the Nernst coefficient \cite{Streda}:
\begin{equation}\label{StredanuA}
\nu^A(T)=\left(\frac{\partial m}{\partial T}\right)_\mu=\left(\frac{\partial s}{\partial B}\right)_\mu,
\end{equation}
where $m$ is magnetization per unit area and $s$ is the entropy per unit area. Eq. (\ref{StredanuA}) and a similar formula for the Hall conductance \cite{Stredasigma} are usually called St\v{r}eda formulas, although they appeared already in \cite{obraztsov}. Eq.~(\ref{StredanuA}) is not exact, but it becomes exact in gapped materials at low $T$ if one assumes that in this limit only surface currents are non-zero \cite{obraztsov,Stredasigma,Streda}. This is discussed further in Appendix A. 

According to the Nernst heat theorem, the entropy $S$ of the cylinder must approach a $B$-independent constant at $T=0$. On the other hand, we can compute the derivative of the entropy with respect to $B$ using (\ref{StredanuA}).
The latter calculation gives
\begin{equation}\label{stredaB}
\left(\frac{\partial S}{\partial B}\right)_\mu=(\nu^A(T)-\nu^A_0(T)) A,
\end{equation}
where $A$ is the area of the cap and $\nu^A_0(T)$ is the Nernst coefficient of the trivial insulator. By the Nernst heat theorem, the l.h.s. of Eq. (\ref{stredaB}) approaches zero as $T\ra 0$. Since the Nernst coefficient of the trivial insulator is zero, this implies $\nu^A(0)=0$. Then Onsager reciprocity implies $\lim_{T\ra 0}\eta^A(T)/T=0$. 

The above argument deserves a few comments. First, we should mention that we have made an implicit assumption, namely that the bulk energy (or mobility) gap remains open for sufficiently small $B$, where $B$ is the magnetic field produced by the solenoid. This assumption is necessary to justify the St\v{r}eda formula, Eq.~(\ref{StredanuA}). 

Our second comment involves an interesting example, namely the $\nu = 5/2$ FQH state. It was proposed by several authors~\cite{CooperStern,YangHalperin} that for this state, the gap $\Delta_n$ for neutral bulk excitations might be much lower than the gap $\Delta_c$ for charged excitations. If $\Delta_n\ll T\ll \Delta_c$, these neutral excitations have an extensive and $B$-dependent entropy, so that $(\partial s/\partial B)_N$ is independent of $T$ and nonzero. Given this $B$ dependent entropy, one might think that the $\nu = 5/2$ FQH state could have a nonzero Nernst coefficient $\nu^A(T)$ at very low temperatures. However, this is \emph{not} the case since the quantity that appears in Eq.~(\ref{stredaB}) is $(\partial s/\partial B)_\mu$ not $(\partial s/\partial B)_N$. Indeed, the results of \cite{CooperStern,YangHalperin} imply that $(\partial s/\partial B)_\mu=0$ in this temperature range, and thus $\nu^A(T)\simeq 0$. 



\section{Equilibrium currents in a unitary CFT}
In this section we show that if edge degrees of freedom of a 2D gapped material are described by a unitary 1+1d CFT with a $U(1)$ symmetry, then the edge $U(1)$ current is independent of $T$. Likewise, we show that the edge energy current has no $\mu$ dependence except for the first term in Eq.~(\ref{eq:edge current 2}). Translating these results into the language of transport coefficients, it follows that $\nu^A(T)$ and $\eta^A(T)/T$ are exponentially suppressed at low temperatures for any material of this kind.

By assumption, the edge excitations are described by a unitary CFT whose operator content includes a traceless symmetric conserved energy-momentum tensor $T$ and a conserved $U(1)$ current $J$. After performing the Wick rotation and introducing complex coordinates \cite{CFTbook}, the operator product expansions of the left-moving energy-momentum $T(z) = \sum_n L_n z^{-n-2}$ and $U(1)$  currents $J(z) = \sum_n J_n z^{-n-1}$ are:
\begin{align}
    T(z) T(w) &\sim \frac {c_L/2}{(z-w)^4} + \frac {2T(w)}{(z-w)^2}+\frac {\partial T(w)}{(z-w)}+{\rm reg},\\\label{TJ OPE}
    T(z) J(w) &\sim \frac {\alpha_L}{(z-w)^3} + \frac {J(w)}{(z-w)^2}+\frac {\partial J(w)}{(z-w)}+{\rm reg},\\
    J(z) J(w) &\sim \frac {k_L}{(z-w)^2}+{\rm reg}.
\end{align}
Analogous expressions are true for the right-moving currents $\bar T$ and $\bar J$. This leads to the following transformation law under a conformal change of coordinates:
\begin{align}
    u'^2 \widetilde T(u) &= T(z) - \frac {c_L} {12} \left\{\frac{u'''}{u'}-\frac{3}{2} \left( \frac {u'' }{u'}\right)^2\right\},\\
    u' \widetilde J(u) &= J(z) - \frac {\alpha_L}{2} \frac{u''}{u'},
\end{align}
where $u(z)$ is a conformal transformation and prime indicates derivatives with respect to $z$. 

Under a gauge transformation $\Omega(z)$ the currents change as follows:
\begin{align}\label{eq:T gauge transformation}
    \widetilde T &= T +\frac {\alpha_L} {2} (\ln \Omega)''+\frac {k_L}2{(\ln \Omega)'} ^2-(\ln \Omega)' J(z),\\
     \widetilde J &= J -  k_L (\ln \Omega)',
\end{align}

The energy and $U(1)$ currents are given by
\begin{align}
    \langle j^E\rangle &= \langle T_{tx} \rangle = \frac 1 {2\pi}\langle T-\bar T\rangle,\\
     \langle j\rangle &= \langle J_{x} \rangle = \frac 1 {2\pi}\langle J-\bar J\rangle,
\end{align}
where $T_{tx}$ is time-space component of the standard energy-momentum tensor and $J_x$ is the charge current. In the plane geometry the expectation values of $j^E$ and $j$ are equal zero due to conformal invariance. Expectation values of the currents at a finite temperature $1/\beta$ can be found by a conformal transformation from a cylinder onto the plane \cite{CFTatFiniteT1,CFTatFiniteT2}:
\begin{equation}
    u(z) = \exp \left[\frac{2\pi i}\beta (\tau+ix)\right], 
\end{equation}
where $z=\tau+ix$ is coordinate on a cylinder with periodic time $\tau \sim \tau + \beta$ and $u$ is a complex coordinate on the plane. Using the transformation law for the currents we find
\begin{align}
    \langle j^E \rangle_{\beta} &= \frac{\pi}{12\beta^2} (c_L-c_R),\\
    \langle j \rangle_{\beta}  &= \frac {i} {2\beta}  (\alpha_L-\alpha_R).
\end{align}
In order to find the currents at non-zero electrochemical potential we can study the behavior of the operators under $U(1)$ gauge transformations. One can see from eq. (\ref{eq:T gauge transformation}) that under a large gauge transformation 
\begin{align}\label{eq:gauge trans}
    \Omega(z)=\exp{(-\mu z)}
\end{align}
the Hamiltonian $H=\dfrac 1 {2\pi} (L_0+\bar L_0)$ transforms as $H\rightarrow H+ \mu Q$ where $Q=\dfrac 1 {2\pi} (J_0+\bar J_0)$. Therefore, a change in the electric potential can be mimicked by the gauge transformation~(\ref{eq:gauge trans}). The currents change as follows:
\begin{align}
    \langle j^E \rangle_{\beta,\mu} &= \frac{\pi}{12\beta^2} (c_L-c_R)+\frac {i\mu} {2\beta} (\alpha_L-\alpha_R)+\frac{(k_L-k_R)}{2\pi}\frac{\mu^2}2\\
    \langle j \rangle_{\beta,\mu}  &= \frac {i} {2\beta}  (\alpha_L-\alpha_R)+\frac{k_L-k_R}{2\pi}\mu .
\end{align}

By considering two edges at slightly different temperatures and chemical potentials  (or by comparing these expressions to (\ref{eq:edge current 1}) and (\ref{eq:edge current 2})) we find the transport coefficients to be
\begin{align}
    \kappa^A &= \frac {\pi}{6\beta}(c_R-c_L),\\
    \nu^A &=\frac {i} {2}  (\alpha_R-\alpha_L),\\
    \eta^A &=\frac {iT} {2}  (\alpha_R-\alpha_L),\\
    \sigma^A&= \frac{k_R-k_L}{2\pi}.
\end{align}

Now we will show that a non-zero value of either $\alpha_L$ or $\alpha_R$ contradicts the unitarity of the CFT. The operator product expansion (\ref{TJ OPE}) leads to the following commutation relation for the modes of the energy-momentum tensor $T(z) = \sum_n L_n z^{-n-2}$ and $U(1)$ current $J(z) = \sum_n J_n z^{-n-1}$:
\begin{align}
    [L_n,J_m] = \alpha_L \frac {(n+1)n} 2 \delta_{n+m,0}-m J_{n+m}.
\end{align}
Using the fact that unitary CFT the has a unique vacuum $|0\rangle$ invariant under global conformal transformations $L_{\pm 1}|0\rangle = L_0|0\rangle =0$, we find
\begin{align}
    0 = \langle [L_1,J_{-1}] \rangle+\langle [L_{-1},J_{1}] \rangle = \alpha_L.
\end{align}
Similarly one can show that $\alpha_R=0$.

\section{Flux insertion argument for vanishing of the Nernst coefficient}
In this section, we present a flux insertion argument showing that the Nernst coefficient $\nu^A(T)$ is exponentially small at low temperatures in any 2D gapped many-body system with short-range interactions and a $U(1)$ symmetry. In the process, we also prove that there is no analog of a Thouless pump for entropy. 

\subsection{Statement of the main result}
Consider a 2D gapped many-body system with short-range interactions and a $U(1)$ symmetry, defined in a \emph{cylinder} geometry. Consider a mixed state of the following form:
\begin{itemize}
\item{The top edge of the cylinder is at temperature $T_t$ and chemical potential $\mu_t$.}
\item{The bottom edge of the cylinder is at temperature $T_b$ and chemical potential $\mu_b$.}
\item{The bulk of the cylinder is in one of a finite set of topologically degenerate ground states.}
\end{itemize}
Here we assume that $T_t, T_b \ll \Delta$ where $\Delta$ is the bulk gap, and that $\mu_t, \mu_b$ are in an appropriate range so that they are consistent with a bulk gap. Physically this mixed state can be prepared by starting the system in one of its ground states, coupling the system to appropriate baths at the two edges, and then decoupling the heat baths from the system.

We will denote the above mixed state by $\rho_0$. More generally, we let $\rho_\theta$ denote the same kind of mixed state as above, but in the presence of magnetic flux $\theta$ through the cylinder, where $0 \leq \theta < 2\pi$. 

A few comments about the mixed states $\rho_0$ and $\rho_\theta$: first, we should mention that $\rho_0$ and $\rho_\theta$ are only \emph{approximately} stationary: if the system is initialized in one of these states, it will eventually relax to a fully equilibrated state in which the two edges are at the same temperatures and chemical potentials. We will mostly neglect this relaxation because it happens at very long time scales: the time scale for the relaxation process is set by dissipative transport coefficients which are exponentially small at low temperatures. Another comment about $\rho_0$ and $\rho_\theta$ is that these mixed states are not uniquely defined in the case where there are multiple topologically degenerate ground states. This ambiguity is not important for our purposes because we will only be interested in local observables, and all the different choices of $\rho_\theta$ share the same expectation values for such observables.

We are now ready to state our main result. Define the ``flux-averaged'' current $\bar{I}$ by
\begin{align}
\bar{I} = \frac{1}{2\pi} \int_0^{2\pi} \mathrm{Tr}(I \rho_\theta) d\theta
\end{align}
where $I$ is the $U(1)$ current operator around the cylinder. Our main result is that $\bar{I}$ is given by
\begin{align}
\bar{I} = \sigma^A(0) (\mu_t -\mu_b) 
\label{I}
\end{align}
up to an error term that is exponentially small for temperatures $T_b, T_t \ll \Delta$. Here, $\sigma^A(0)$ denotes the zero temperature Hall conductance of the gapped many-body system. 

We can go a step further if we make the ``flux-averaging assumption'' that $\mathrm{Tr}(I \rho_\theta)$ is independent of $\theta$. Under that assumption, Eq.~(\ref{I}) implies that
 \begin{align}
I = \sigma^A(0) (\mu_t -\mu_b) 
\label{I2}
\end{align}
where $I$ is the expectation value of the current for any \emph{fixed} value of flux, say $\theta = 0$. 

The most important implication of these results is that $\bar{I}$ and $I$ do not depend on the temperature of the top or bottom edge, except for terms that are exponentially small for temperatures $T_t, T_b \ll \Delta$. This lack of temperature dependence means that the Nernst coefficient $\nu^A(T) = \frac{dI}{dT}$ is also exponentially small for temperatures $T \ll \Delta$.

\subsection{Outline of the argument}
Our argument is based on a flux insertion process similar to that of Laughlin~\cite{Laughlin_flux}. We imagine initializing the system in the (zero flux) mixed state $\rho_0$ described above. We then imagine slowly inserting $2\pi$ flux through the hole of the cylinder. Here when we say ``slowly'' we mean that the flux should be inserted over a time scale $\mathcal{T}$ that is much longer than $1/\Delta$ and also much longer than $\tau$ where $\tau$ is the relaxation time scale associated with the edge excitations. We will also assume that the flux insertion time scale $\mathcal{T}$ is much \emph{shorter} than the exponentially long time scale associated with equilibration between the two edges. This hierarchy of time scales is important because it guarantees that the flux insertion process is a ``quasi-static'' process -- i.e. each edge remains in local thermal equilibrium throughout the process.

We make two claims about this flux insertion experiment which we will prove below. Our first claim is a finite temperature variant of one of the standard claims from Laughlin's original flux insertion argument~\cite{Laughlin_flux}:
\begin{claim}
\label{claim1}
The following identity holds:
\begin{align}
\bar{I} = \frac{\Delta E}{2\pi} 
\label{claim1eq}
\end{align}
where $\Delta E$ is the change in the expectation value of the total energy of the cylinder when $2\pi$ flux is inserted. 
\end{claim}

Our second claim is less familiar but can be derived from a basic thermodynamic inequality, together with locality properties of the flux insertion process:
\begin{claim}
\label{claim2}
The following inequalities hold:
\begin{align}
\Delta E_t \geq \mu_t \Delta N_t, \quad \quad \Delta E_b \geq \mu_b \Delta N_b
\label{claim2eq}
\end{align}
where $\Delta E_t$ and $\Delta N_t$ are the changes in the expectation values of the energy and number of particles near the top edge when $2 \pi$ flux is inserted through the cylinder, and  $\Delta E_b$ and $\Delta N_b$ are defined similarly, but near the bottom edge.
\end{claim}

Once we prove these claims, we can easily derive our main result, Eq.~(\ref{I}). The first step is to note that 
\begin{align}
\Delta E = \Delta E_t + \Delta E_b
\label{EtEb}
\end{align}
since the flux insertion process does not change the energy density in the bulk (i.e. it returns the bulk to one of its ground states). Next, we note that the quantities $\Delta N_t$ and $\Delta N_b$ are related to the zero temperature Hall conductance $\sigma^A(0)$ by
\begin{align}
\Delta N_t = -\Delta N_b = 2\pi \sigma^A(0)
\label{sigmaH}
\end{align}
up to exponentially small corrections. Then, we combine (\ref{claim1eq}), (\ref{claim2eq}), (\ref{EtEb}) and (\ref{sigmaH}) to deduce the inequality
\begin{align}
\bar{I} \geq \sigma^A(0) (\mu_t - \mu_b)
\label{Iineq1}
\end{align}
Next, imagine rotating the cylinder by 180 degrees (exchanging the top and bottom of the cylinder). This operation changes $\mu_b \leftrightarrow \mu_t$, and replaces $\bar{I} \rightarrow -\bar{I}$, while preserving the Hall conductance $\sigma^A(0)$, so we deduce the inequality
\begin{align}
-\bar{I} \geq \sigma^A(0) (\mu_b - \mu_t)
\label{Iineq2}
\end{align}
Combining the two inequalities (\ref{Iineq1}), (\ref{Iineq2}) proves the result (\ref{I}).

In the next two sections we give physical arguments for Claims~\ref{claim1} and \ref{claim2}. 

\subsection{Physical argument for Claim~\ref{claim1}}

To prove Claim~\ref{claim1}, we directly compute the change in the expectation value of the energy of the cylinder, $\Delta E$. 

First, we need to introduce some notation. Let $H$ denote the initial Hamiltonian and let $\theta(t)$ denote the flux through the cylinder at time $t$. We define the corresponding time dependent Hamiltonian $H(t)$ as follows: we choose a branch cut that runs from one end of the cylinder to the other, and then we ``twist'' all the terms in $H$ that straddle this branch cut by conjugating them by the unitary operator $e^{i \theta(t) Q_+}$ where $Q_+$ denotes the total $U(1)$ charge on one side of the branch cut. A convenient feature of this gauge choice is that the initial and final Hamiltonians are the same, i.e. $H(0) = H(\mathcal{T})$, since $\theta(\mathcal{T}) = 2\pi$.

Next, let $U(t)$ denote the unitary time evolution operator, 
\begin{align}
U(t) = \mathbf{T} \exp \left[ -i \int_0^t dt' H(t') \right].
\end{align}
Finally, let $U \equiv U(\mathcal{T})$ denote the time evolution operator for the whole flux insertion process.

With this notation the change in the expectation value of the energy is given by
\begin{align}
\Delta E &= \mathrm{Tr}[H  U \rho_0 U^\dagger] - \mathrm{Tr}(H \rho_0)
\end{align}
Rewriting this expression as an integral over $t$ gives 
\begin{align}
\Delta E &= \int_0^{\mathcal{T}} \frac{d}{dt} \mathrm{Tr}[H(t) U(t) \rho_0 U^\dagger(t)] dt \nonumber \\
&= \int_0^{\mathcal{T}} \mathrm{Tr}\left[\frac{dH}{dt} U(t) \rho_0 U^\dagger(t) \right] dt  \nonumber \\ 
&= \int_0^{\mathcal{T}} \mathrm{Tr}\left[\frac{\partial H}{\partial \theta}U(t) \rho_0 U^\dagger(t) \right] \frac{d\theta}{dt} dt  \nonumber \\
&= \int_0^{\mathcal{T}} \mathrm{Tr}\left[I U(t) \rho_0 U^\dagger(t) \right] \frac{d\theta}{dt} dt
\label{de1} 
\end{align}
Here, the third equality follows from the fact that the time dependence of $H$ comes entirely from the time dependent flux $\theta(t)$, while the last equality follows from the fact that $I = \frac{\partial H}{\partial \theta}$.

So far everything is exact, but to proceed further we need to invoke physical arguments. The first step is to note that since the flux insertion process is \emph{quasistatic}, the density matrix at time $t$, namely $U(t) \rho_0 U^\dagger(t)$, shares approximately the same expectation values for local operators as a (local) equilibrium density matrix $\rho_{eq}(t)$ of the following form: $\rho_{eq}(t)$ describes a state where the top of the cylinder is at temperature $T_t(t)$ and chemical potential $\mu_t(t)$ and the bottom of the cylinder is at temperature $T_b(t)$ and chemical potential $\mu_b(t)$, and where there is flux $\theta(t)$ through the hole of the cylinder. In other words,
\begin{align}
 \mathrm{Tr}\left[\mathcal{O} U(t) \rho_0 U^\dagger(t) \right] \approx  \mathrm{Tr}\left[\mathcal{O} \rho_{eq}(t)\right]
\label{Oid1}
\end{align}
for any local operator $\mathcal{O}$. Here the ``$\approx$'' sign means that the error vanishes in the thermodynamic limit. 

The next step is to note that the time-dependent temperatures and chemical potentials $T_t(t), \mu_t(t), T_b(t), \mu_b(t)$ that appear in $\rho_{eq}(t)$ only differ from their initial values $T_t, \mu_t, T_b, \mu_b$ by an amount that vanishes in the thermodynamic limit. To see this, note that the flux insertion process can only change the energy/number of particles on a given edge by a quantity of at most order $O(1)$, so it cannot affect the temperature or chemical potential of either edge when we take the thermodynamic limit. This means that we can replace $\rho_{eq}(t) \rightarrow \rho_{\theta(t)}$ when computing expectation values: i.e.,
\begin{align}
 \mathrm{Tr}\left[\mathcal{O}  \rho_{eq}(t) \right] \approx  \mathrm{Tr}\left[\mathcal{O} \rho_{\theta(t)}\right]
\label{Oid2}
\end{align}
for any local operator $\mathcal{O}$. Again, the ``$\approx$'' sign means that the error vanishes in the thermodynamic limit. 

Combining (\ref{Oid1}) and (\ref{Oid2}), and using $\mathcal{O} = I$, we derive
\begin{align}
 \mathrm{Tr}\left[I U(t) \rho_0 U^\dagger(t) \right] \approx \mathrm{Tr}\left[I \rho_{\theta(t)} \right]
\label{Iid}
\end{align}
where the error vanishes in the thermodynamic limit.\footnote{Readers may object that $I$ is not a local operator, but rather a sum of local operators along a branch cut, and hence Eq.~(\ref{Iid}) does not follow. However the crucial point is that $I$ is a local operator in the \emph{circumferential} direction. This locality in the circumferential direction is all that we need to justify (\ref{Iid}).}

With Eq.~(\ref{Iid}) in hand, the rest of the derivation follows from straightforward algebra. Substituting (\ref{Iid}) into Eq.~(\ref{de1}), we derive
\begin{align}
\Delta E &= \int_0^{\mathcal{T}} \mathrm{Tr}\left[I \rho_{\theta(t)} \right] \frac{d\theta}{dt} dt  \nonumber \\
&= \int_0^{2\pi} \mathrm{Tr}\left[I \rho_\theta \right] d\theta \nonumber \\
&= 2\pi \bar{I}
\label{calc1}
\end{align}
This completes our proof of Claim~\ref{claim1}.

\subsection{Physical argument for Claim~\ref{claim2}}

We now give a physical argument for Claim~\ref{claim2}. Our argument is based on two properties of the flux insertion process: (i) the flux insertion process does not create bulk excitations, and (ii) the flux insertion process takes a finite amount of time that does not scale with the length of the cylinder: that is, the unitary $U$ that implements the flux insertion process can be written in the form
\begin{align}
U = \mathbf{T} \exp \left(-i \int_0^\mathcal{T} H(t) dt \right)
\label{localunit2}
\end{align}
where $H(t)$ is a local Hamiltonian, and $\mathcal{T}$ does not scale with the length of the cylinder. 

In order to explain the argument we need to introduce some notation for labeling the low energy edge excitations of the cylinder (in the absence of flux): we label these states as $|i,j,a\>$ where $i$ labels the edge states at the bottom of the cylinder, $j$ labels the edge states at the top, and $a$ labels the topological sector of the system. Note that, despite the simple notation, each state $|i,j,a\>$ is generally a complicated and highly entangled many-body wave function.

The topological sector $a$ will not play an important role below, since we will assume that the cylinder is initialized in a single topological sector, and furthermore we will assume that the flux insertion process does not change the topological sector.\footnote{We can guarantee the latter property by inserting $2\pi m$ flux instead of $2\pi$ flux, and taking $m$ to be a multiple of $1/e^*$ where $e^*$ is the smallest fractionally charged excitation.} Thus, the system will always be in the same sector throughout our discussion. For this reason, we will drop the ``$a$'' index from now on and denote the low energy states by $|i,j\>$.

Next, we need to discuss the \emph{quantum numbers} associated with each eigenstate $|i,j\>$. Because the two edges are well-separated, we assume that the energy of $|i,j\>$ can be written as a sum of the form $E_i^b + E_j^t$ for some real constants $\{E_i^b\}, \{E_j^t\}$, i.e.
\begin{align}
H |i,j\> = (E_i^b + E_j^t)|i,j\>
\end{align}
Likewise, we assume that the total particle number of $|i,j\>$ is of the form $N_i^b + N_j^t$, i.e.
\begin{align}
N |i,j\> = (N_i^b + N_j^t)|i,j\>
\end{align}

With this notation, we can write down an explicit formula for the initial density matrix of the cylinder, $\rho_0$:
\begin{align}
\rho_0 &= \sum_{ii'jj'} \rho_{ii'}^b \rho_{jj'}^t |i,j\>\<i',j'|, \nonumber \\
\rho_{ii'}^b &= \frac{1}{Z_b} e^{-(E_i^b - \mu_b N_i^b)/T_b} \delta_{ii'}, \quad \quad \rho_{jj'}^t = \frac{1}{Z_t} e^{-(E_j^t - \mu_t N_j^t)/T_t} \delta_{jj'}
\end{align}

Next, consider the \emph{final} density matrix, $\rho_f = U \rho_0 U^\dagger$. Since the flux insertion process does not introduce any bulk excitations we know that $\rho_f$ must be of the form
\begin{align}
\rho_f = \sum_{ii'jj'} A_{ii'jj'} |i,j\>\<i',j'|
\end{align}
for some coefficients $A_{ii'jj'}$. In fact, we can say more: using the fact that $U$ is of the form given in Eq.~(\ref{localunit2}) where $\mathcal{T}$ does not scale with the length of the cylinder, it is possible to show that $A_{ii' jj'}$ can be factored as
\begin{align}
A_{ii'jj'} = \sigma^b_{ii'} \sigma^t_{jj'}
\label{Afact1}
\end{align}
where $\sigma^b_{ii'}$ and $\sigma^t_{jj'}$ are Hermitian matrices with the same eigenvalue spectrum as $\rho^b_{ii'}$ and $\rho^t_{jj'}$:
\begin{align}
\mathrm{Spec}(\sigma^b) &= \mathrm{Spec}(\rho^b) \nonumber \\
\mathrm{Spec}(\sigma^t) &= \mathrm{Spec}(\rho^t)
\label{sigmarhoV}
\end{align}
We give the proof of Eqs.~(\ref{Afact1}), (\ref{sigmarhoV}) in Appendix~\ref{proofapp}. 

To proceed further, we use the following result, which is a restatement of the well-known fact that the Gibbs state minimizes the free energy $F = E - TS$: 

\begin{lemma} 
\label{lemma1}
Let $H$ be a Hermitian matrix, and let $\bar\rho$ be a matrix of the form
\begin{align}
\bar{\rho} = \frac{1}{Z} e^{-H/T}, \quad \quad Z = \mathrm{Tr}(e^{-H/T})
\end{align}
for some non-negative real $T$. Let $\rho$ be another matrix of the same dimension as $\rho$ such that $\rho$ is positive semi-definite and $\mathrm{Tr}(\rho) = 1$. Then
\begin{align}
\mathrm{Tr}(H \rho + T \rho \log \rho)  \geq \mathrm{Tr}(H \bar{\rho} + T \bar{\rho} \log \bar{\rho})
\end{align}
\end{lemma} 
This inequality can be derived straightforwardly by minimizing the convex functional $F[\rho] = \mathrm{Tr}(H \rho + T \rho \log \rho) $.

First we apply Lemma~\ref{lemma1} with $\rho = \sigma^t$, and $\bar{\rho} = \rho^t$ and with $H$ being the diagonal matrix $(E_i^t  - \mu_t N_i^t )\delta_{ii'}$. This gives the inequality
\begin{align}
\sum_{i} (E_i^t - \mu_t N_i^t) \sigma^t_{ii} + T_t \cdot \mathrm{Tr}(\sigma^t \log \sigma^t) \geq \sum_{i} (E_i^t - \mu_t N_i^t) \rho^t_{ii} + T_t \cdot \mathrm{Tr}(\rho^t \log \rho^t)
\end{align}

Next, invoking Eq.~(\ref{sigmarhoV}), we can cancel the $\mathrm{Tr}(\sigma^t \log \sigma^t)$ and  $\mathrm{Tr}(\rho^t \log \rho^t)$ terms on the two sides to obtain
\begin{align}
\sum_{i} (E_i^t - \mu_t N_i^t) \sigma^t_{ii} \geq \sum_{i} (E_i^t - \mu_t N_i^t) \rho^t_{ii} 
\end{align}
Subtracting the right hand side from the left hand side gives the inequality
\begin{align}
\Delta E_t - \mu_t \cdot \Delta N_t \geq 0
\end{align}
where $\Delta E_t$, $\Delta N_t$ denote the change in the expectation value of the energy and particle number at the top edge during the flux insertion process. 

In the same way, we can apply Lemma 1 with $\rho = \sigma^b$, and $\bar{\rho} = \rho^b$  and with $H$ being the diagonal matrix $(E_i^b  - \mu_b N_i^b )\delta_{ii'}$ to derive
\begin{align}
\Delta E_b - \mu_b \cdot \Delta N_b \geq 0
\end{align}
This proves Claim~\ref{claim2}.

\subsection{Impossibility of a Thouless pump for entropy}

Using the thermodynamic identity, $\Delta E = T \Delta S + \mu \Delta N$, we can identify the two quantities $\Delta E_t - \mu_t \Delta N_t$ and $\Delta E_b - \mu_b \Delta N_b$ in the statement of Claim~\ref{claim2} with $T_t \Delta S_t$ and $T_b \Delta S_b$ where $\Delta S_t$ and $\Delta S_b$ are the change in entropy at the top and bottom edges. With these identifications, Claim~\ref{claim2} implies that 
\begin{align}
\Delta S_t \geq 0, \quad \quad \Delta S_b \geq 0. 
\end{align}

One implication of the above inequalities is that they rule out the possibility that the flux insertion process could pump entropy from one end of the cylinder to the other, i.e. the possibility that $\Delta S_t = - \Delta S_b \neq 0$. In fact, we can go a step further: since the proof of Claim~\ref{claim2} does not use any of the details of the flux insertion process, we can rule out the possibility of \emph{any} adiabatic cycle consisting of local, quasi-1D Hamiltonians $H(\theta)$ with a bulk energy gap that pumps a nonzero amount of entropy across the system at temperatures below the bulk gap. In other words, we deduce that it is impossible to construct an analog of the (1D) Thouless pump for entropy.

We note that a weaker\footnote{The second version of the no-go result is weaker than the first because it only shows that $\lim_{T \rightarrow 0} \Delta S = 0$, while the first argument implies that $\Delta S$ is exponentially small for $T$ smaller than the bulk gap.} version of this no-go result can be derived directly from the Nernst unattainability principle. Specifically, the Nernst principle implies that for any adiabatic cycle $H(\theta)$ of the above type, the amount of entropy $\Delta S$ pumped across the system must vanish as $T \rightarrow 0$, i.e. $\lim_{T \rightarrow 0} \Delta S = 0$. To show this, we use the same argument in Sec.~\ref{sec:thirdlaw}: if $\Delta S$ remained nonzero in this limit, then we could use this adiabatic cycle to cool a finite heat bath to zero temperature in a finite number of cycles, which would contradict the Nernst unattainability principle.

\section{Concluding remarks}

It is often stated that the Third Law of Thermodynamics requires the tensors $\nu_{km}(T)$ and $\eta_{km}(T)/T$ to vanish at $T=0$, see e.g. \cite{CRHu}. The discussion in this paper shows that the relation between the Third Law and the behavior of thermoelectric coefficients near $T=0$ is rather subtle. On the one hand, the Nernst unattainability principle directly implies the vanishing of the symmetric tensor $\eta^S(T)/T$ at $T=0$. Applying this  both to the original system and its time-reversal and using Onsager reciprocity, one concludes that  $\nu^S(0)=0$ as well. On the other hand, a non-vanishing value of $\eta^A(T)/T$ at $T=0$ is associated only with circulating entropy currents and does not directly conflict with the Nernst unattainability principle. For gapped systems, however, one can use the St\v{r}eda formula to show that a nonzero value for $\nu^A(0)$ is in conflict with the Nernst heat theorem (which follows from the Nernst unattainability principle and some standard assumptions). Then Onsager reciprocity implies the vanishing of $\eta^A(T)/T$ at $T=0$ as well. 

Among our arguments for the vanishing of $\nu^A(0)$, the one based on the Third Law of Thermodynamics is the most robust, but also the least informative. It shows that $\nu^A(0)=0$ but does not tells us anything about the magnitude of $\nu^A(T)$ for small but nonzero $T$. The other arguments (Sections IV and V) rely on more assumptions but show that $\nu^A(T)$ is exponentially small for $T$ below the bulk gap. 

One may ask for an intuitive reason why the thermal Hall effect gives rise to a topological invariant of gapped systems, while the Nernst and Ettinghausen effects do not. Using the same line of reasoning as in Appendix A, one can show that the edge contribution to the thermal Hall conductivity is given by a Str\v{e}da-like formula 
\begin{equation}
\kappa^A_{edge}=\frac{\partial m^E}{\partial T}.
\end{equation}
Here $m^E$ is the ``energy magnetization'' per unit area defined by the equation for the equilibrium energy current ${\bf j}^E_k=\eps_{kl} \partial_l m^E.$ For a gapped 2d system one expects the bulk contribution to $\kappa^A$ to be exponentially suppressed at low $T$. However, unlike in the case of the Nernst coefficient, there is no Maxwell relation which would allow one to express $\kappa^A$ as the  derivative of the entropy density with respect to a parameter. This is because there is no gravitational analog of the magnetic field which could contribute a term $m^E dB^E$ to the variation of the free energy density. 
Furthermore, even if one could express $\kappa^A$ as a derivative of the entropy density, such a formula would only show that $\kappa^A(T)$ vanishes at $T=0$ and would not constrain the \emph{slope} of $\kappa^A(T)$ at low $T$ -- the actual topological invariant in the case of the thermal Hall effect~\cite{thermalHallpaper}. 

\newpage
\begin{appendix}
\section{St\v{r}eda formulas}
In this appendix we recall the derivation of St\v{r}eda formulas for the Hall and Nernst coefficients. The argument is essentially the same as the one presented in \cite{Streda}.

Consider a finite size homogeneous system coupled to two reservoirs with temperatures $T_1,T_2$ and electrochemical potentials $\mu_1,\mu_2$ respectively. The system will relax to a non-equilibrium steady state with a linearly changing temperature $T({\bf r})$ and electrochemical potential $\mu({\bf r})$. In the bulk of the sample away from the edges and reservoirs, the electric current can be decomposed as  
\begin{multline}
   j_i({\bf r }) =    j_i^{\rm tr}({\bf r })+j_i^{\rm mag}({\bf r })=- \sigma_{ik} \partial_k \mu - \nu_{ik} \partial_k T+ \eps_{ik} \partial_k m({\bf r})\\=- \sigma_{ik} \partial_k \mu- \nu_{ik} \partial_k T + \eps_{ik} \partial_k \mu({\bf r}) \frac { \partial m({\bf r})}{ \partial \mu}+ \eps_{ik} \partial_k T({\bf r}) \frac { \partial m({\bf r})}{ \partial T},
\end{multline}
where $j_i^{\rm mag}({\bf r })= \epsilon_{ik} \partial_k m({\bf r})$ is circulating currents which don't contribute to the net current across any section, and $j_i^{\rm tr}({\bf r })= - \sigma_{ik} \partial_k \mu- \nu_{ik} \partial_k T$ is the transport current. In the gapped phase at low temperature all bulk current across any macroscopic section should be exponentially suppressed for any values of $\partial_k \mu$ and $\partial_k T$. Therefore we find  
\begin{align}
     \sigma_{ik}&= \epsilon_{ik} \frac { \partial m}{ \partial \mu},\\
     \nu_{ik}&= \epsilon_{ik} \frac { \partial m}{ \partial T}.
\end{align}
One can see from the derivation that these formulas are only correct up to exponentially suppressed terms in the  temperature arising from bulk currents.
On the other hand, 
it can be shown that St\v{r}eda formulas provide exact expressions for ``static'' Hall and Nernst coefficients which describe the equilibrium response of a system to an electric field and a temperature gradient \cite{auerbach}.

\section{Proving Eqs.~(\ref{Afact1}), (\ref{sigmarhoV})}
\label{proofapp}
We now prove Eqs.~(\ref{Afact1}), (\ref{sigmarhoV}). To this end, we need to introduce some notation. First, we let $\mathcal{V}_a$ denote the subspace spanned by all edge states $|i,j,a\>$ within a fixed topological sector $a$. As in the main text we will drop the ``a'' index from now on, since it always be fixed. Thus we will use the notation $\mathcal{V}_a \rightarrow \mathcal{V}$ and $|i,j,a\> \rightarrow |i,j\>$ in what follows.

Next, we let ``$|\Omega, \Omega\>$'' denote the state $|i,j\>$ in $\mathcal{V}$ with the minimum value of the ``energy'' 
\begin{align}
(E^b_i - \mu_b N^b_i) + (E^t_j - \mu_t N^t_j). 
\end{align}
We will assume for simplicity that 
\begin{align}
E_\Omega^b - \mu_b N^b_\Omega = 0, \quad \quad E_\Omega^t - \mu_t N^t_\Omega = 0, 
\end{align}
so that $|\Omega, \Omega\>$ has an ``energy'' of exactly $0$.

A final piece of notation: for each $\delta > 0$, we define $\mathcal{V}_\delta$ to be subspace of $\mathcal{V}$ spanned by states of the form $|i,j\>$ with
\begin{align}
E_i^b - \mu_b N^b_i \leq \delta, \quad \quad E_j^t - \mu_t N^t_j \leq \delta
\end{align}
Roughly speaking, $\mathcal{V}_\delta$ contains all states with ``energies'' of at most $\delta$ on both edges.

Our proof relies on three assumptions about the $|i,j\>$ edge states:
\begin{itemize}
\item{{\bf Assumption 1: (Existence of local operators)} For each $i,i'$, there exists an operator $O_{i'i}^b$ supported near the bottom edge such that
$O_{i'i}^b |i,j\> = |i',j\>$ for all $j$. Likewise, for each $j,j'$ there exists an operator $O_{j'j}^t$ supported near the top edge such that $O_{j'j} |i,j\> = |i,j'\>$ for all $i$.}

\item{{\bf Assumption 2: (Short range correlations)} For each $i,j$, the state $|i,j\>$ has short-range correlations between the top and bottom edge: that is, for any operators $O^b, O^t$ supported near the bottom and top edges respectively, 
\begin{align}
\<i,j| O^b O^t |i,j\> = \<i,j| O^b |i,j\> \<i,j| O^t |i,j\>
\end{align}
}

\item{{\bf Assumption 3: ($U$ does not create bulk excitations)} There exists a $\delta > 0$ such that $U \cdot \mathcal{V}_\delta \subset \mathcal{V}$. We will also assume that $T_b, T_t \ll \delta$ so that $\rho_0$ is a mixture of states $|i,j\>$ belonging to $\mathcal{V}_\delta$.
}
\end{itemize}
We will now use these assumptions to prove the claims. Our argument proceeds in three steps.

\subsection{Step 1: Showing that $U |\Omega, \Omega\>$ is factorizable}
In the first step, we will show that there exists operators $S_b, S_t$ support near the bottom and top edges respectively, such that 
\begin{align}
U |\Omega, \Omega\> = S^b S^t |\Omega, \Omega\>
\label{sbst}
\end{align}
To prove this, we note that Assumption 3 implies that $U |\Omega, \Omega\> \in \mathcal{V}$. Hence, 
\begin{align}
U |\Omega, \Omega\> = \sum_{ij} X_{ij} |i,j\>
\label{Uomega}
\end{align}
for some complex coefficients $X_{ij}$. Multiplying this equation by its adjoint, we conclude that
\begin{align}
U |\Omega, \Omega\> \<\Omega, \Omega| U^\dagger = \sum_{ii'jj'} X_{ij} X^{*}_{i'j'} |i,j\>\<i',j'|
\end{align}
or equivalently
\begin{align}
U \rho_\Omega U^\dagger = \sum_{ii'jj'} X_{ij} X^{*}_{i'j'} |i,j\>\<i',j'|
\end{align}
where $\rho_\Omega \equiv |\Omega, \Omega\> \<\Omega, \Omega|$. Using Assumption 1, we deduce that
\begin{align}
\text{Tr}( O_{i'i}^b O_{j'j}^t U \rho_\Omega U^\dagger) = X_{ij} X^{*}_{i'j'}
\end{align}
We can rewrite this relation as
\begin{align}
\text{Tr}( \tilde{O}_{i'i}^b \tilde{O}_{j'j}^t \rho_\Omega) = X_{ij} X^{*}_{i'j'}
\label{Omegarel1}
\end{align}
where
\begin{align}
\tilde{O}^b_{i'i} = U^\dagger O^b_{i'i} U, \quad \quad \tilde{O}^t_{j'j} = U^\dagger O^b_{j'j} U
\label{tildenot}
\end{align}
Now, since $U$ is a local unitary transformation of the form (\ref{localunit2}), we know that $\tilde{O}^b_{i'i}$ and
$\tilde{O}^t_{j'j}$ are supported near the bottom and top edges respectively (this follows from Lieb-Robinson bounds~\cite{hastings2010locality}). Therefore, by Assumption 2, we can factor the left hand side of (\ref{Omegarel1}) to obtain
\begin{align}
\text{Tr}( \tilde{O}_{i'i}^b \rho_\Omega) \cdot \text{Tr}(\tilde{O}_{j'j}^t \rho_\Omega) = X_{ij} X^{*}_{i'j'}
\label{Omegarel2}
\end{align}
Equivalently, we can write this relation as
\begin{align}
\tau^b_{i' i} \tau^t_{j' j} =  X_{ij} X^{*}_{i'j'}
\label{Omegarel3}
\end{align}
where
\begin{align}
\tau^b_{i'i} = \text{Tr}( \tilde{O}_{i'i}^b \rho_\Omega) , \quad \quad \tau^t_{j'j} = \text{Tr}(\tilde{O}_{j'j}^t \rho_\Omega)
\end{align}

An immediate mathematical consequence of (\ref{Omegarel3}) is that $X_{ij}$ can be factored as
\begin{align}
X_{ij} = \alpha_i \beta_j
\label{Xab}
\end{align}
for some complex coefficients $\alpha_i,\beta_j$. One way to see this is to note that the right hand side of (\ref{Omegarel3}) looks like the density matrix for a pure state with wave function $X_{ij}$. From this point of view, Eq.~(\ref{Omegarel3}) implies that the density matrix corresponding to $X_{ij}$ can be written as the tensor product of two density matrices for $i$ and $j$ separately. Hence $X_{ij}$ has no entanglement between $i$ and $j$, which implies that $X_{ij}$ can be written in the form (\ref{Xab}).

We now substitute (\ref{Xab}) into (\ref{Uomega}) to derive
\begin{align}
U |\Omega, \Omega\> = \sum_{ij} \alpha_i \beta_j |i,j\>
\label{Uomega2}
\end{align}
We can now construct the required operators $S^b, S^t$:
\begin{align}
S^b = \sum_i \alpha_i O^b_{i \Omega}, \quad \quad S^t = \sum_j \beta_j O^t_{j \Omega}
\end{align}
By construction, $S^b, S^t$ obey equation (\ref{sbst}). This proves the claim.

\subsection{Step 2: Showing $\<i, j| U |i',j'\>$ is factorizable}
In the second step, we show that  $\<i, j| U |i',j'\>$ is factorizable, i.e.
\begin{align}
\<i, j| U |i',j'\> = Y^b_{ii'} Y^t_{jj'}
\label{Ufactor}
\end{align}
for some complex coefficients $Y^b_{ii'}, Y^t_{jj'}$.

To show this, we note that by Assumption 1, we can write $\<i, j| U |i',j'\>$ as
\begin{align}
\<i, j| U |i',j'\> = \<\Omega, \Omega| (O^t_{j \Omega})^\dagger (O^b_{i \Omega})^\dagger U O^b_{i' \Omega} O^t_{j' \Omega} |\Omega, \Omega\> 
\end{align}
Equivalently, we can write this as
\begin{align}
\<i, j| U |i',j'\> = \<\Omega, \Omega| (O^t_{j \Omega})^\dagger (O^b_{i \Omega})^\dagger  \bar{O}^b_{i' \Omega} \bar{O}^t_{j' \Omega} U |\Omega, \Omega\> 
\end{align}
where $\bar{O}^b_{i'i} = U {O}^b_{i'i} U^\dagger$, and $ \bar{O}^t_{j'j} = U O^t_{j'j} U^\dagger$.

Next, using (\ref{sbst}), we can rewrite this as
\begin{align}
\<i, j| U |i',j'\> = \<\Omega, \Omega| (O^t_{j \Omega})^\dagger (O^b_{i \Omega})^\dagger  \bar{O}^b_{i' \Omega} \bar{O}^t_{j' \Omega} S^b S^t |\Omega, \Omega\> 
\end{align}

Now using Assumption 2, we can factor the right hand side into two parts:
\begin{align*}
\<i, j| U |i',j'\> = Y^b_{ii'} Y^t_{jj'}
\end{align*}
where
\begin{align}
Y^b_{ii'} = \<\Omega, \Omega| (O^b_{i \Omega})^\dagger \bar{O}^b_{i' \Omega} S^b |\Omega, \Omega\> \nonumber \\
Y^t_{jj'} = \<\Omega, \Omega| (O^t_{j \Omega})^\dagger \bar{O}^t_{j' \Omega} S^t |\Omega, \Omega\> \nonumber \\
\end{align}
This establishes the factorization (\ref{Ufactor}).

\subsection{Finishing the proof}
We are now ready to finish the proof: we will prove Eqs.~(\ref{Afact1}), (\ref{sigmarhoV}). The first step is to note that the matrices $Y^b$ and $Y^t$ that appear in (\ref{Ufactor}) are guaranteed to obey a unitarity property. To be precise, $Y^b$ and $Y^t$ obey
\begin{align}
[(Y^b)^\dagger Y^b]_{i'i} = \delta_{i'i}, \quad \quad [(Y^t)^\dagger Y^t]_{j'j} = \delta_{j'j}
\label{uniprop}
\end{align}
for any $i,i',j,j'$ such that $|i,j\>, |i',j'\> \in \mathcal{V}_\delta$ where $\mathcal{V}_\delta$ is defined as in Assumption 3 above. Indeed, the above property follows from the fact that $U$ is unitary and that $U|i,j\>, U|i',j'\> \in V$ (according to Assumption 3).\footnote{More precisely, the fact that $U$ is unitary implies that $Y^b, Y^t$ can always be rescaled by a scalar factor $Y^b \rightarrow Y^b \cdot \lambda$,  $Y^t \rightarrow Y^t \cdot \lambda^{-1}$ so that they obey the unitarity property (\ref{uniprop}). We will assume this rescaling in what follows.}

To prove (\ref{Afact1}), we note that the initial density matrix $\rho_0$ can be factorized as
\begin{align}
\<i,j| \rho_0 |i',j'\> = \rho^b_{ii'} \rho^t_{jj'}
\end{align}
Therefore, by (\ref{Ufactor}), we have
\begin{align}
\<i, j|U \rho_0 U^\dagger |i', j'\> =  \sigma^b_{ii'} \sigma^t_{jj'}
\end{align}
where
\begin{align}
\sigma^b = Y^b \rho^b (Y^b)^\dagger, \quad \quad \sigma^t = Y^t \rho^t (Y^t)^\dagger
\end{align}
Furthermore, it is clear from the above expressions that $\sigma^b$ and $\sigma^t$ are Hermitian. This completes our proof of (\ref{Afact1}).

As for (\ref{sigmarhoV}), the unitarity property (\ref{uniprop}) implies that
\begin{align}
\text{Tr}[(\sigma^b)^n] = \text{Tr}[(\rho^b)^n] \nonumber \\
\text{Tr}[(\sigma^t)^n] = \text{Tr}[(\rho^t)^n]
\end{align}
for any positive integer $n$ and hence
\begin{align}
\text{Spec}(\sigma^b) = \text{Spec}(\rho^b) \nonumber \\
\text{Spec}(\sigma^t) = \text{Spec}(\rho^t)
\end{align}
This completes our proof of (\ref{sigmarhoV}).

\end{appendix}

\bibliographystyle{apsrev4-1}
\bibliography{bib}

\end{document}